\documentclass[aps,twocolumn,showlabels,showrefs,amsmath,amssymb,pre,superscriptaddress, floatfix, colors]{revtex4}

\usepackage{graphicx}
\usepackage{dcolumn}
\usepackage{bm}
\usepackage{amssymb}
\usepackage{multirow}
\usepackage{color}
\usepackage{hyperref}

\begin{document}

\title{Collective motion of Active Brownian Particles with polar alignment}

\author{Aitor Mart\'in-G\'omez} 
\email{aitormg93@gmail.com}
\affiliation{Theoretical Soft Matter and Biophysics, Institute of Complex Systems and Institute for Advanced Simulation, Forschungszentrum J\"ulich, D-52425 J\"ulich, Germany}

\author{Demian Levis} 
\affiliation{CECAM Centre Europ\'een de Calcul Atomique et Mol\'eculaire, \'Ecole Polytechnique F\'ed\'erale de Lausanne, Batochimie, Avenue Forel 2, 1015 Lausanne, Switzerland}
\affiliation{University of Barcelona Institute of Complex Systems (UBICS), Universitat de
Barcelona, Barcelona, Spain}

\author{Albert D\'iaz-Guilera} 
\affiliation{University of Barcelona Institute of Complex Systems (UBICS), Universitat de Barcelona, Barcelona, Spain}
\affiliation{Departament de F\'isica de la Mat\`eria Condensada, Universitat de Barcelona, Mart\'i i Franqu\`es, E08028 Barcelona, Spain}

\author{Ignacio Pagonabarraga} 
\affiliation{CECAM Centre Europ\'een de Calcul Atomique et Mol\'eculaire, \'Ecole Polytechnique F\'ed\'erale de Lausanne, Batochimie, Avenue Forel 2, 1015 Lausanne, Switzerland}
\affiliation{University of Barcelona Institute of Complex Systems (UBICS), Universitat de
Barcelona, Barcelona, Spain}
\affiliation{Departament de F\'isica de la Mat\`eria Condensada, Universitat de Barcelona, Mart\'i i Franqu\`es, E08028 Barcelona, Spain}

\begin{abstract}
We present a comprehensive computational study of the collective behavior emerging from the competition between self-propulsion, excluded volume interactions and velocity-alignment in a two-dimensionnal model of active particles. We consider an extension of the  Active Brownian Particles model where the self-propulsion direction of the particles aligns with the one of their neighbors.   
We analyze the onset of collective motion (flocking) in a low-density regime ($10\% $ surface area) and show that it is mainly controlled by the strength of velocity-alignment interactions: the competition between self-propulsion and crowding effects plays a minor role in the emergence of flocking. 
However, above the flocking threshold, 
 the system presents a richer pattern formation scenario than analogous models without alignment interactions (Active Brownian Particles) or excluded volume effects (Vicsek-like models).
Depending on the parameter regime, the structure of the system is characterized by either a broad distribution of finite-sized polar clusters or the presence of an amorphous, highly fluctuating, large-scale traveling structure which can take a lane-like or band-like form (and usually a hybrid structure which is halfway in between both). 
We establish a phase diagram that summarizes collective behavior of polar Active Brownian Particles and propose a generic mechanism to describe the complexity of the large-scale structures observed in systems of repulsive self-propelled particles.   
\end{abstract}

\pacs{pacs}
\maketitle

\setlength{\textfloatsep}{10pt} 
\setlength{\intextsep}{10pt}

 
\section{Introduction}\label{sec:Introduction}

Flocks of birds, bacterial colonies, synthetic diffusio-phoretic colloidal suspensions or dense assemblies of actin filaments are all examples of active matter systems, where the individual constituents, or 'active particles', pump energy from the environment to convert it into directed motion. In the presence of interactions, this ability to self-propel leads to interesting out-of-equilibrium many-body phenomena, 
such as the emergence of collective motion (or flocking), observed in populations of active particles  (both in living and man-made systems) across a broad range of scales: from vertebrates down to motile bacteria or intracellular filaments driven by molecular motors \cite{VicsekRev, BechingerRev}. 
From a general, fundamental perspective, the emergence of collective  motion is understood as arising  from the presence of velocity-alignment interactions between self-propelled 'point-like' agents, as put forward by the celebrated Vicsek model \cite{Vicsek1995} and its fore-coming refinements and extensions \cite{VicsekRev, Toner1995, Toner2005, Ramaswamy2010}. 
A salient feature of this non-equilibrium symmetry breaking phase transition, is the spontaneous formation of large-scale structures in the form of traveling bands \cite{Chate2008, Solon2015}: the emergence of order in Vicsek models is followed by spatio-temporal heterogeneities.

The ability to generate self-sustained structures constitutes a generic, distinctive feature of active systems. Indeed, the formation of structures, or patterns, driven by activity is not always associated to the emergence of collective motion. 
Isotropic self-propelled particles, with no alignment interactions, show a tendency to aggregate into dense clusters \cite{Theurkauff2012, Fily2012, Levis2014, Soto2014, Liebchen2015}, eventually triggering a  macroscopically phase separation 
in the absence of attractive forces. This so-called Motility-Induced Phase Separation (MIPS) is mainly understood as arising from the mere competition between excluded volume interactions and self-propulsion, as discussed extensively in the context of Active Brownian Particle (ABP) models \cite{CatesRev, Fily2012, Redner2013, Stenhammar2013, Stenhammar2014, Levis2017}. 

However, any experimental realization involves more intricate effects and interactions than purely velocity alignment rules or excluded volume. Consistently, experiments usually show more complex structures, or patterns, probably due to the interplay between different interaction mechanisms. Despite the ubiquity of excluded volume interactions, which should be present in any system of polar active particles - whether is wet, e.g. suspensions of micro-swimmers, or dry, e.g. animal flocks \cite{MarchettiRev} - its impact on the flocking transition and its corresponding patterns remains unclear. 
Several studies have shown that excluded volume interactions might considerably affect the formation of structures in these systems, leading to band-like patterns coexisting with lanes, polarized clusters, asters and vortices \cite{Schaller2010, Schaller2010, Schaller2011polar, Kohler2011, Farrell2012, Bricard2015}. Despite some works dealing with excluded-volume effects \cite{Szabo2006, Grossman2008, Henkes2011, Lam2015},  whether  aligning rigid-particle systems display the same pattern-forming mechanisms and critical properties as the original Vicsek model of point-like agents  or not remains an open issue.  
From the theory side, a continuum formulation of a smooth version of the Vicsek-like model with density dependent self-propulsion velocity was proposed in \cite{Farrell2012}. This approach, motivated by previous work on MIPS,  aims to capture particle interactions in an effective way, amenable to a continuum formulation. 
While the onset of flocking is unchanged by the density-dependent velocity effective interactions (sometimes called 'quorum sensing'), above it,  in the  ordered phase, interactions induce  richer structure formation as compared to Vicsek-like models of point-like particles. 

Here, we aim at characterizing the emergence of collective motion and out-of-equilibrium patterns in a system of aligning self-propelled particles with short-range repulsions. 
We introduce a model of polar self-propelled particles build as an extension of the paradigmatic Active Brownian Particles and Vicsek models. 
This approach allows us to disentangle the role played by velocity-alignment and volume exclusion interactions, bridging together the aggregation phenomena  discussed in the context of isotropic and polar active particles.
After the definition of the model in section \ref{sec:Model}, we perform a systematic computational study of its phase behavior. In section \ref{sec:PhD} we summarize our findings in a phase diagram. We first describe the emergence of flocking in the system in section  \ref{sec:flock}, to then move to the characterization of the patterns  in section  \ref{sec:cluster}. Finally,  section \ref{sec:conclusion} is devoted to a general discussion of our results.


\section{Model}\label{sec:Model}

The system consists on a two-dimensional assembly of $N$ Active Brownian Particles (ABP) in a square box of size $L_x\times L_y$, subjected to periodic boundary conditions (PBC). The dynamics of the system is governed by the over-damped  stochastic equations:
\begin{equation}
\dot{\mathbf{r}}_i (t) = v_0\mathbf{n}_i(t) + \mu\mathbf{F}_i \, ,
\label{r_dot}
\end{equation}
\begin{equation}
\dot \theta _i (t) = \frac{K}{\pi R_{\theta}^2} \sum_{j \in \partial_i} \sin(\theta_j - \theta_i) + \sqrt{\gamma}\eta_i(t) \, .
\label{theta_dot}
\end{equation}
Here $\mathbf{r}_i(t) = (x_i,y_i)$ denotes the spatial position of the $i$-th particle at time $t$,  $v_0$ the constant magnitude of the self-propulsion velocity along its orientation $\mathbf{n}_i(t)=(\cos \theta_i,\, \sin \theta_i)$, $\mu$ is the mobility and $\mathbf{F}_i$ is the inter-particle repulsion force accounting for excluded volume between particles which derives from a continuous potential 
$$\mathbf{F}_i= - \nabla_i \sum_{j < i} v(r_{ij}=|\mathbf{r}_i-\mathbf{r}_j|),\, v(r)= \epsilon \left(\frac{\sigma}{r}\right)^{12}\, ,$$
with an upper cutoff at $3\sigma$.  The parameter $\sigma$ can be thought of as the effective particle diameter. From now on $\sigma=1$, fixing the unit of length. 
The term $\eta_i$ is a zero-mean white noise with unit-variance. It accounts for fluctuations in the self-propulsion direction, whose strength is controlled by the parameter $\gamma$, 
introducing time correlations between displacements over a typical timescale $\tau=\gamma^{-1}$, hence called the \emph{persistence time}.
The first term in eq. \eqref{theta_dot} is a continuous-time version of a Vicsek-like interaction: it controls the velocity alignment between the particles, leading to local orientational ordering with a coupling strength $K$ (normalized by the interaction area, see below). This coupling strength defines the characteristic time-scale associated to the phase dynamics $\tau_K=\pi R_{\theta}^2/K$. Here, in the spirit of previous modeling of ABP with alignment interactions \cite{Peruani2008mean, Chepizhko2010, Farrell2012,  CAP, CAPhard}, we introduce a ferromagnetic-like coupling between  velocities akin to the one of the Kuramoto model of phase synchronization \cite{KuramotoRev} or, equivalently, the XY model in two dimensions. The sum $\sum_{j \in \partial_i}$ runs over all the particles $j$ in the vicinity of particle $i$, i.e. particles which  verify  $r_{ij} \leq R_{\theta}$.  $R_{\theta}$ thus defines the coupling range of the alignment interaction which is held fixed $R_{\theta}=2\sigma$ (This parameter has to be chosen larger than $\sigma$ in order to guarantee the presence of alignment interactions in the infinitely hard disk limit).   

It is useful at this stage to identify the relevant non-dimensional parameters that control the collective behavior of the system. 
From the Langevin equation \eqref{r_dot}-\eqref{theta_dot} we  identify:
$$\text{Pe}= \frac{v_0}{\sigma \gamma}, $$
the rotational P\'eclet number, which quantifies the self-propulsion strength of the particles, or, equivalently, their persistence length $l_p=\sigma\text{Pe}$;
$$\text{g}=\frac{K}{4 \pi \sigma^2 \gamma},$$
the dimensionless coupling parameter,  which compares the coupling strength, i.e. the tendency of the particles to align their velocities, with the angular noise intensity that tends to randomly reorient them; and $$\Gamma= \frac{\mu \epsilon}{\sigma v_0 }\, ,$$
providing a measure of the effective particle stiffness as the self-propulsion is varied. 
In the present work, we restrict ourselves to the regime where excluded volume interactions always dominate  over  self-propulsion and the particles can be effectively described as  hard disks of diameter $\sigma$ (note that our repulsive potential diverges at the origin, thus guaranteeing the above mentioned condition). 
Finally, the relevance of crowding is controlled by the effective packing fraction 
$$\phi=\frac{\pi N}{4L_xL_y} \, .$$ 
The collective behavior of the system is purely controlled by these three dimensionless parameters: $\phi$,  $\text{Pe}$ and $\text{g}$. 
In this work we focus on the phase behavior of Aligning Active Brownian Particles at fixed $\phi=0.1$ and investigate in detail the role played by velocity alignment and excluded volume interactions. 


In the weak coupling limit, $\text{g} \to 0$, we recover the (isotropic) 
ABP model, which has become over the last years the prototypical model to study the competition between steric effects and self-propulsion. 
As it has been largely discussed in the  literature \cite{CatesRev, Fily2012, Redner2013, Stenhammar2014, Levis2017}, at high enough density and activity, the combination of excluded volume interactions and self-propulsion provokes a Motility-Induced Phase Separation (MIPS). Here, we analyze systems which lie below the critical packing fraction and P\'eclet number needed in order to nucleate a macroscopic cluster from an homogenous state and observe MIPS. 
In this regime, ABP show a clear tendency to form clusters that do not coarsen above a certain size \cite{Fily2012, Levis2014,Levis2017}. 

In the limit of point-like particles ($\epsilon\to 0$), our model reduces to a system of self-propelled particles with velocity alignment akin to the one proposed by Vicsek \cite{Vicsek1995} but in a continuous-time formulation \cite{Ratushnaya2007, Farrell2012, Chepizhko2013, CAP, CAPhard}.
As such, our model should reproduce in this limit the main collective features of Vicsek-like models.  At low values of $\text{g}$ the system sets into a  disordered homogeneous state. For values above the flocking threshold, $\text{g}>\text{g}^c$, the system develops global polar order. Flocking is understood as a first-order-like phase transition, characterized by a coexistence region where high density bands coexist with a disordered low-density background. For higher values of the coupling strength, the system eventually leaves the coexistence region, rendering the traveling bands unstable and leading to the so-called fluctuating flocking state or Toner-Tu phase \cite{Toner1995, Toner2005, Ramaswamy2010, Chate2008, Mishra2010}. 


The aligning term, Eq. (\ref{theta_dot}), is reminiscent of the (noisy) Kuramoto model of synchronization phenomena and the planar XY model of magnetism. Indeed, our model can be thought of as a Kuramoto model on a time evolving network generated by the positions of the particles \cite{AlbertPRX}. The internal variable $\theta_i$, is  equivalent to the phase of an oscillator which here dictates its self-propulsion direction.  In the absence of motion, i.e. $v_0\to 0$ the model reduces to a system of identical oscillators in a static network governed by a  Kuramoto dynamics with a tendency towards phase synchronization or, equivalently, to the XY model on the 2D geometric network generated by the position of each particle, connected which each other as soon as they are distant of $r<R_{\theta}$.   
In this static limit, it is well known that the Mermin-Wagner theorem forbids the development of long-range order in 2D. The role of mobility in this context has been investigated recently \cite{Uriu2013,  Grossmann2015, AlbertPRX}. 
For a network which evolves in time as the result of the self-propelled motion of its nodes, the 2D XY universality class is preserved  as soon as the 'internal' state of the XY-spin attached to each node is decoupled to its dynamics in real space \cite{AlbertPRX}. Hence, a mobile system cannot sustain long range order if  the density field and the polarization field are decoupled. 
However, once the 2D spin (or oscillator phase) dictates the direction of self-propulsion, i.e. the density field is advected by the polarization field, the situation changes qualitatively:  the system can then sustain long-range polar order as predicted by the Toner-Tu phase and  featuring  giant density fluctuations. 

 \begin{figure*}
  \includegraphics[scale=0.53]{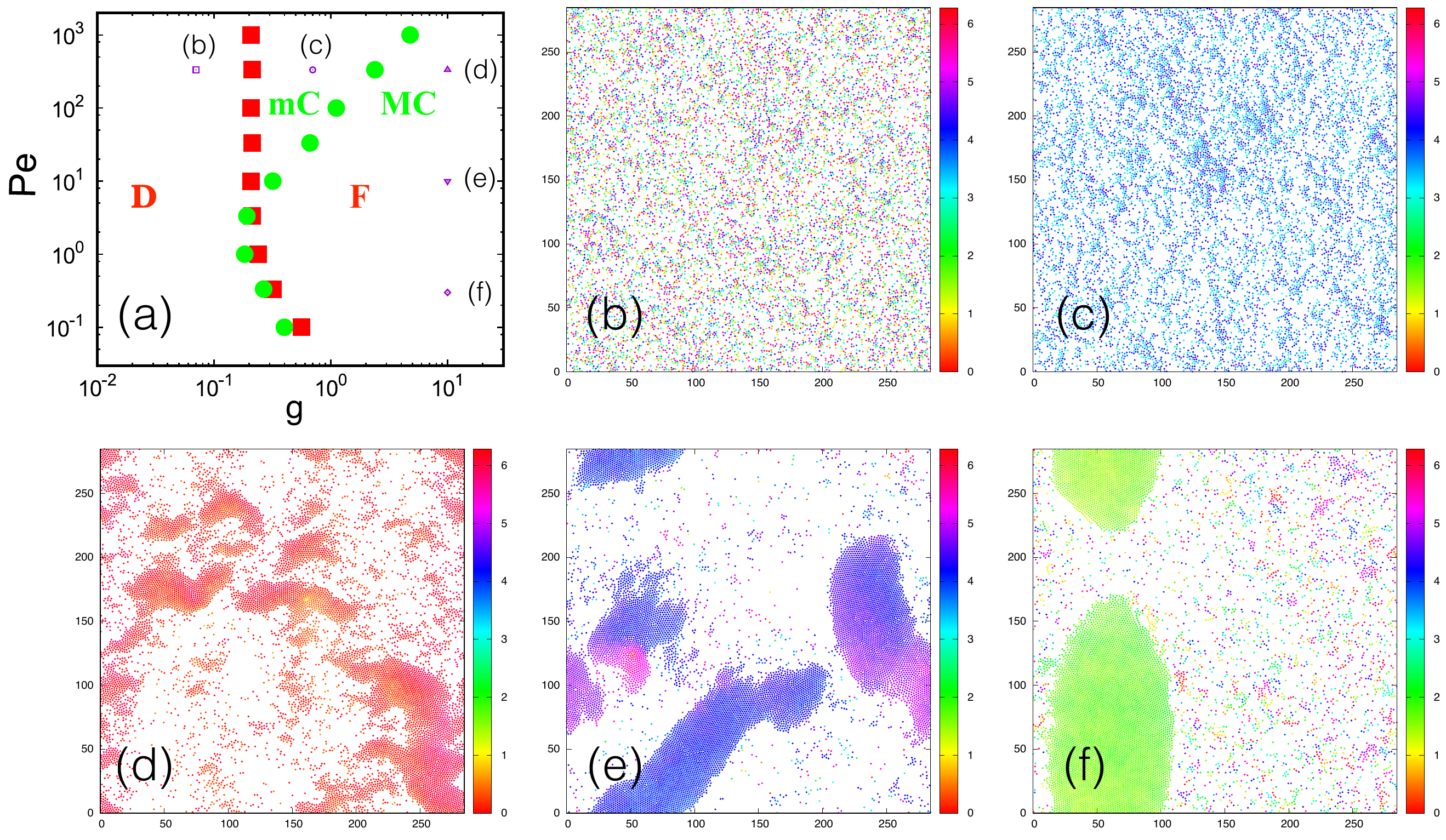} 
\caption{(a) Phase diagram in the $\text{g}$-$\text{Pe}$ plane. The red symbols correspond to the onset of flocking   $\text{g}^c$. We represent by green symbols our estimation of the structural transition  $\text{g}^*$ characterized by the first appearance of a macroscopic dense structure. (b-f) Representative snapshots of a system of $N=10^4$ corresponding to the values of the parameters labelled in panel (a):   ($\text{Pe}$ , $\text{g}$) = (333.3 , 0.07); (333.3 , 0.7); (333.3 , 10); (10 , 10); (0.3 , 10).  The color code represents the orientation $\theta_i$ of each particle.  }\label{fig:PhD}
\end{figure*}

In order to study the phase behavior of the model in the $(\mbox{Pe},\, g)$-plane, we integrate the over-damped Langevin equations (\ref{r_dot}), (\ref{theta_dot}) using the Euler method with a time step in the range $\delta t=10^{-3}-10^{-1}$ (depending on the value of the parameters). 
The white noise term $\eta_i$ eq. (\ref{theta_dot}) is drawn from a flat distribution in the interval $[-1, 1]$ independently for each particle at each time-step. 
The simulations have been performed for different noise intensities, $\gamma$, and coupling strengths, $K$, while keeping $v_0=0.1$ and  $\epsilon=1$ fixed. 
We explore easily up to four orders of magnitude in the $(\mbox{Pe},\, g)$-plane: $\text{Pe}=[10^{-1},10^{3}]$ and $\text{g}=[10^{-2},10^{1}]$ by varying $\gamma=10^{-4},...,1$ and  $K=10^{-4},...,10$.
In order to identify the relevance of  finite-size corrections, we have explored systems of $N=1000$ up to $10000$ particles with periodic boundary conditions (PBC) both in square $L_x=L_y=L$ and slab $L_x=5L_y$ geometries. We use a slab geometry in order to properly identify bands and lanes, as explained below (see section \ref{sec:cluster}). 

\section{Phase behaviour}\label{sec:PhD}

We start our analysis by first discussing the steady states of the model for systems defined on a square $L\times L$ box with PBC. Our systematic analysis resulted in the phase diagram shown in Fig.\ref{fig:PhD} (a).

For small $\text{g}$, below the critical value  $\text{g}^c$ (shown by red symbols in Fig.\ref{fig:PhD} (a)), we find the system in a disordered state (D), followed by a flocking phase (F) at $\text{g}>\text{g}^c$, characterized by global polar order induced by the alignment interactions, as expected from the literature of flocking \cite{VicsekRev}. In simple Vicsek-like models of point-like particles, the flocking state is characterized by the emergence of high density bands close enough to the onset of flocking   \cite{Chate2008,Solon2015}.    
Besides this tendency to develop long-range polar order and band patterns due to velocity alignment, in our model, particles aggregate into isotropic clusters as a result of the competition between self-propulsion and short-range repulsions. 
The interplay between these two interactions, gives rise to a different mechanism for structure formation that we analyze here. 

Below the critical value $\text{g}^c$ the system is  a non-polar fluid state populated by a distribution of clusters that grow with $\text{Pe}$. This regime is controlled by the physics of ABP (see Fig.\ref{fig:PhD} (b)).   
However, once rotational symmetry has been broken the situation changes radically and new structures, reminiscent to those reported in \cite{Farrell2012}, emerge.  
At a given finite value of $\text{Pe}$ and for  $\text{g}>\text{g}^c$, the model exhibits a phase transition from an isotropic state of zero polarization, to an ordered  state characterized by the collective motion of the system, which acquires a net polarization, $P>0$, and the presence strong density heterogeneities.
The alignment interaction polarizes the clusters and promotes the formation of structures whose size increases with $\text{g}$, eventually leading to a (macroscopic) large high density structure (see the snapshots Fig.\ref{fig:PhD} (d-f)). 

Indeed, for a given $\text{Pe}$, there is a critical alignment $\text{g}^*$ above which a macroscopic polar cluster appears (hence called MC phase) . The location $\text{g}^*$ is indicated by green symbols in the phase diagram  Fig.\ref{fig:PhD} (a). 
Instead, in the intermediate region between the flocking threshold and the emergence of a macroscopic structure, $\text{g}^c<\text{g}<\text{g}^*$, the system sets in a state made of a distribution of coherently moving finite-size `microscopic'  polar clusters  (hence called mC phase). As shown in Fig. \ref{fig:CDF}, as we increase $\text{g}$ at fixed $\text{Pe}$ the cluster size distribution broadens and a transition towards a  state characterized by the appearance of a `macroscopic' cluster  (MC) is observed at $\text{g}^*$. 

In the low persistence regime  $\text{Pe}\lesssim 1$ the advent of a macroscopic polar structure is controlled by the establishment of  polar order in the system, such that $\text{g}^*\approx \text{g}^c$. 
Interestingly, as the persistence of the particles increases, the flocking and structural transitions decouple, leaving a wide area in the phase diagram where global polar order does not induce the formation of a macroscopic structure, as opposed to the liquid-gas picture of the flocking transition in Vicsek-like models with constant velocity \cite{Solon2015}. This suggests that, while correlated, a different mechanism governs the emergence of density heterogeneities and global polar order in our case. This translates into a richer scenario for structure formation than in apolar ABP systems  or point-like self-propelled polar particles. 

For large alignment, in the MC phase,  the system is characterized  by the presence of a macroscopic cluster. This structure can either be a \emph{band}- or \emph{lane}-like traveling structure (and usually a fluctuating mixture of both structures is observed). The difference between both, relies on the relative motion of the particles that form the structure with respect to the band geometry: for band-like structures, particles move orthogonally to the long-axis of the cluster (see Fig. \ref{fig:PhD} (d)), as a propagating front, while for lanes,  they move parallel to it  (see Fig.\ref{fig:PhD} (f)). While bands are generically found in systems of aligning self-propelled point-like particles, lanes have been observed in more realistic systems where crowding effects are at play \cite{Farrell2012, Schaller2010, Schaller2011polar, Kohler2011}. 
In our model, transversal (band) and longitudinal (lane) swarming patterns coexist in the MC regime.
As we discuss below, we provide evidences that the pattern-selection process is controlled by the persistence of the particles, and propose a mechanism to explain why lanes generically appear in systems with excluded volume interactions.  At high persistence and alignment, the heterogeneous structure of the system is dominated by the presence of bands (see Fig. \ref{fig:PhD} (d)). By decreasing $\text{Pe}$,  band structures become rarer while lanes continuously take over to finally dominate the low-$\text{Pe}$ regime (see Fig. \ref{fig:PhD} (e,f) and supplementary movies \cite{SM}). In order to favor the formation of such structures and analyze their relative statistical weight as as a function of $\text{Pe}$, we simulate the model in a rectangular box   $L_x=5L_y$. We postpone this discussion to  section \ref{sec:cluster}.

\section{Flocking}\label{sec:flock}

\begin{figure}
\centering
{\includegraphics[scale=0.85]{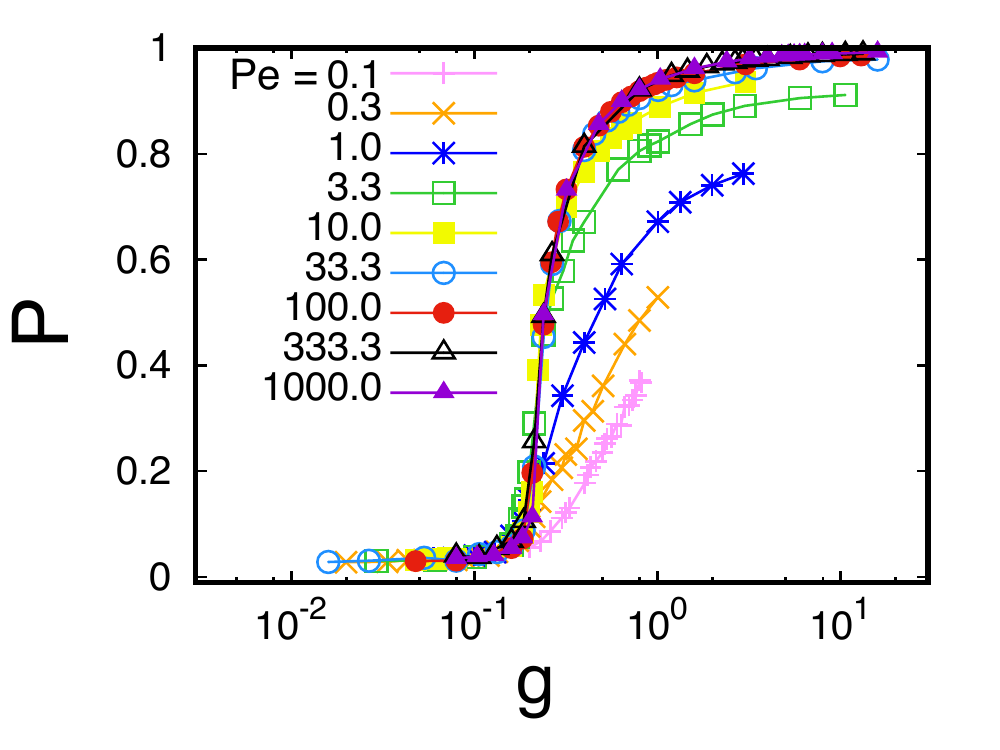}}
{\includegraphics[scale=0.85]{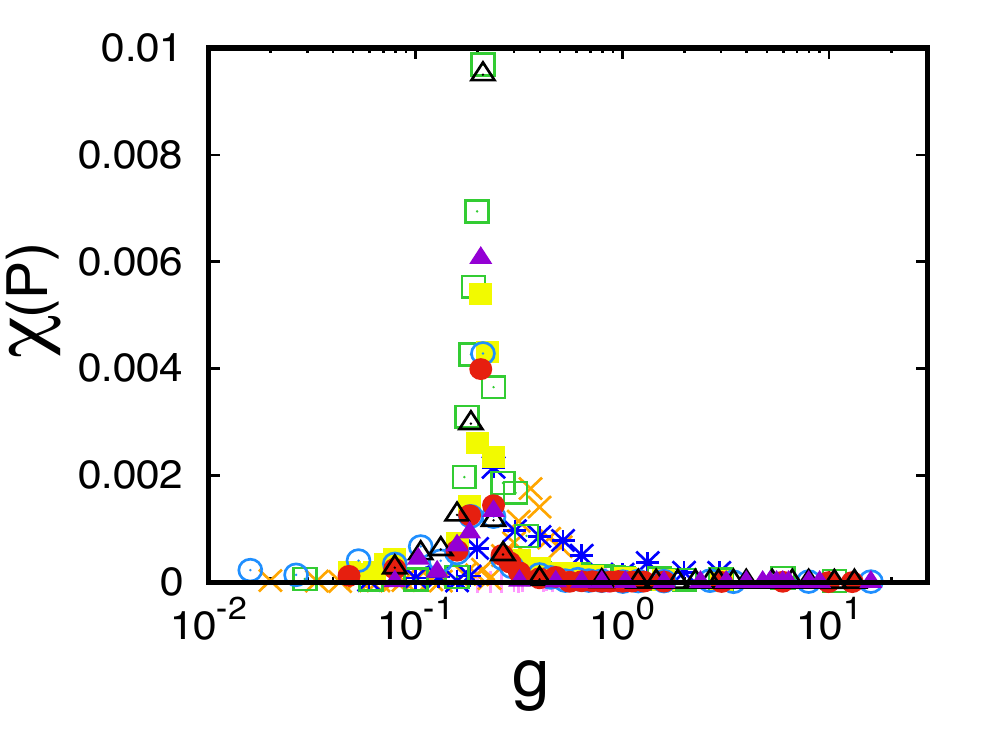}}
\caption{Global polarization as function of $\text{g}$ for several values of $\text{Pe}$ shown in the key (top panel) and its associated susceptibility (bottom panel). 
}\label{fig:Zchi}
\end{figure}

We focus now on the emergence of spontaneous polar order (flocking) as the normalized coupling strength  $\text{g}$ increases. In this section we provide the details about the location (and nature) of the flocking transition  (in red in Fig. \ref{fig:PhD} (a)). 

We characterize the amount of  order in our system by the global polarization
\begin{equation}
P=\left\| \frac{1}{N}\sum_i^N \mathbf{n}_i\right\| \, .
\label{op}
\end{equation}
as typically defined in the context of Vicsek-like models. 
In Fig.\ref{fig:Zchi} (a)  we show the behavior of $P$ as a function of $\text{g}$, for several  $\text{Pe}$ and averaged over 10-100 independent configurations. 
As it is commonly done in the study of equilibrium phase transitions, we analyze the susceptibility of the order parameter $P$ defined as
\begin{equation}
\chi= N \left( \langle P^2 \rangle - \langle P\rangle^2\right) \, ,
\end{equation} 
where $\langle \cdot \rangle$ represents  an average over independent realizations. 
As shown in Fig. \ref{fig:Zchi} (b), $\chi$ is peaked at a value close to the onset of flocking expected from the polarization data.  The location of the flocking transition is weakly dependent on $\text{Pe}$, in agreement with  continuum hydrodynamic theories \cite{Farrell2012, CAP}. The presence of this peak allows us to locate the transition point: we identify the flocking transition (represented by red symbol in the phase diagram Fig. \ref{fig:PhD} (a)) with the maximum of $\chi$.  Our data shows that the location of the flocking transition is, up to numerical accuracy, independent of the particle's persistence  for $\text{Pe}>1$ and around $\text{g}^c=0.21 \pm 0.02$. 


Nonetheless, from Fig. \ref{fig:Zchi}, we notice that the net polarization drops gradually and the polar transition shifts towards higher  $\text{g}$ as we decrease  the particles persistence down to $\text{Pe}<1$. 
The limit $\text{Pe}\to0$ can be approached by taking $v_0\to0$ at fixed $\gamma$ or, conversely,  $\gamma\to\infty$ at fixed $v_0$. 
In the first case, $v_0\to0$, particles generate a Wigner-like crystal, because of the absence of translational noise and the repulsive nature of the interaction potential. Moreover, at the density considered here, the system cannot find a globally oriented state, even in the limit $\text{g}\to\infty$, since the interaction network is not simply connected (a geometrical percolation  occurs above $R_{\theta}\approx \sqrt{4.51/\rho\pi}$ \cite{Dall2002}$\approx3.6\sigma$ here). We shall note that, even at higher densities the system would not be able to sustain a net polarization, since the Mermin-Wagner theorem forbids any 2D static network to globally order. The emergence ofglobal order in models of flocking is purely driven by self-propulsion.  
In the limit of diverging noise intensity $\gamma\to\infty$, which is the way we approach here the limit $\text{Pe}\to0$,  the alignment interaction has to be increasingly large to compete with the noise term eq. \ref{theta_dot} and the motion of the particles becomes basically Brownian.  In such a large noise amplitude regime, an agreement with the mean field theory prediction  \cite{Farrell2012}, stating that the flocking threshold does not depend upon $\text{Pe}$, should not  be expected.

\section{Clustering}\label{sec:cluster}

\begin{figure}
  \includegraphics[scale=0.35]{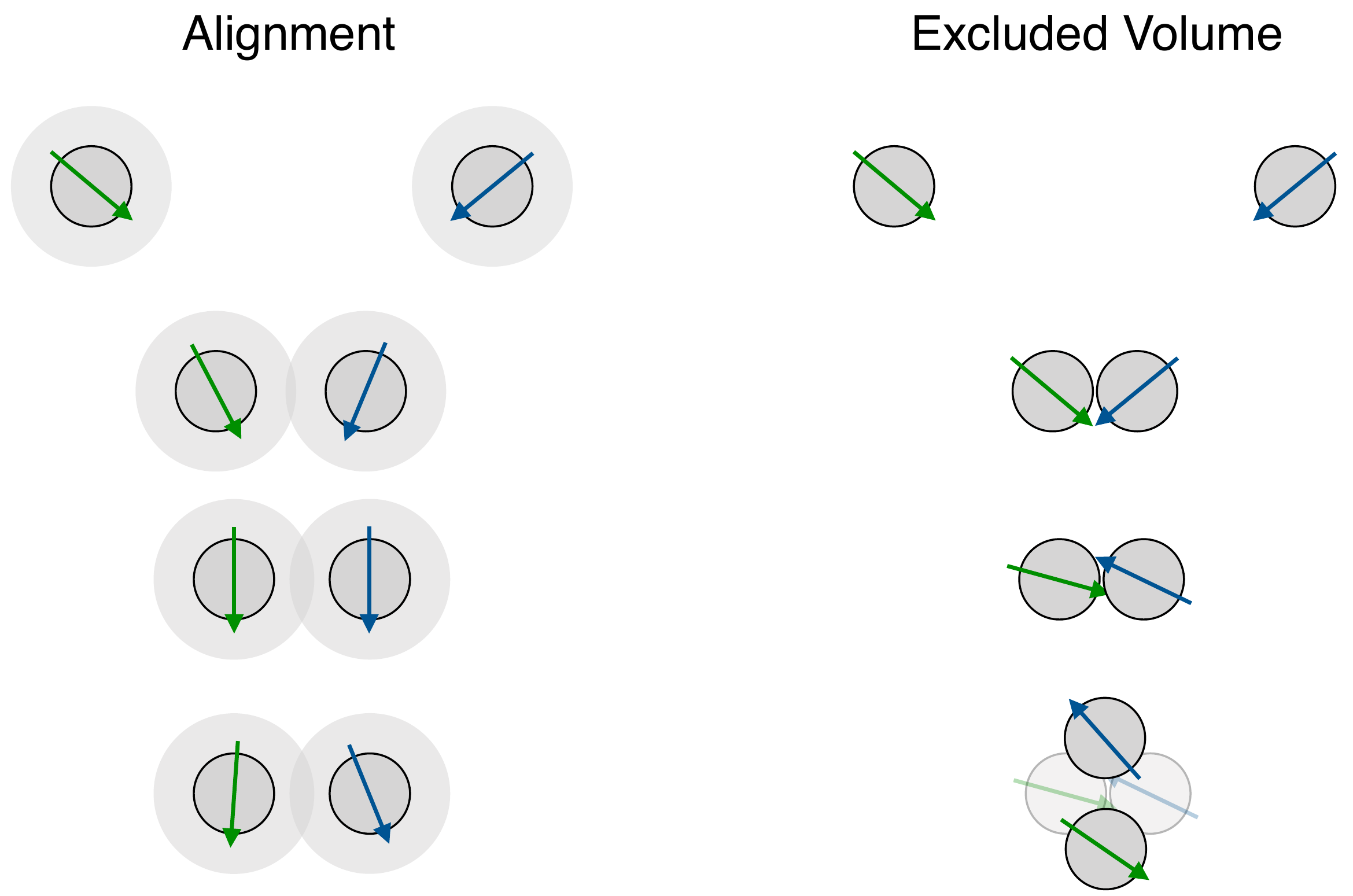} 
\caption{Illustration of the clustering mechanism due to alignment interactions (left) and active brownian motion in the presence of excluded volume interactions (right). The grey shaded area around the particles in the left panel represents the alignment-interaction area (of radius $R_{\theta}=2\sigma$). 
}\label{aline}
\end{figure}


After the description of the phase behaviour of the model in the $(\mbox{Pe},\,g)$-plane, we turn now into the characterization of the different self-sustained structures emerging in the system, in particular in the MC regime, where band- and lane-like patterns coexist. 

As it has now been widely reported,  active particles show a tendency to aggregate in the absence of  inter-particle attractive interactions. Clustering in our model comes from the natural competition between self-propulsion, alignment interaction and excluded volume. As illustrated in Fig. \ref{aline},  two different aggregation mechanisms are involved: a first one driven by  velocity alignment, as in Vicsek models, and a second one dominated by the self-trapping of particles because of the persistence of their motion.  

Let us focus first on the high coupling regime ($\text{g} \gg 1$). In this regime, when two particles are close-by, they align their direction of motion in a shorter time scale than the one associated to the reorientation of their self-propulsion velocity. As a result, particles are kinetically trapped in the interaction area, at a typical distance of the order of $R_{\theta}$, and move along (on average) parallel trajectories for long periods of times. As the density increases the collision rate  grows, such that subsequent particles accumulate in the neighborhood more easily, growing a cluster which, in turn, can capture more particles.  Then, only a rare random fluctuation of the orientation of a particle in the cluster's boundary can make it escape.  

In the absence of alignment ($\text{g} \ll 1$), self-propelled disks slow down as they collide. They hinder their relative motion for a period of time of the order of $\tau$, until their orientation changes by rotational diffusion. At high enough densities, this self-trapping mechanism, illustrated in Fig. \ref{aline}, is at the origin of MIPS in ABP systems. While this mechanism is present in our model, as we show, it is not the leading one in the parameter regimes we explored. 

\begin{figure}
  \includegraphics[scale=0.8]{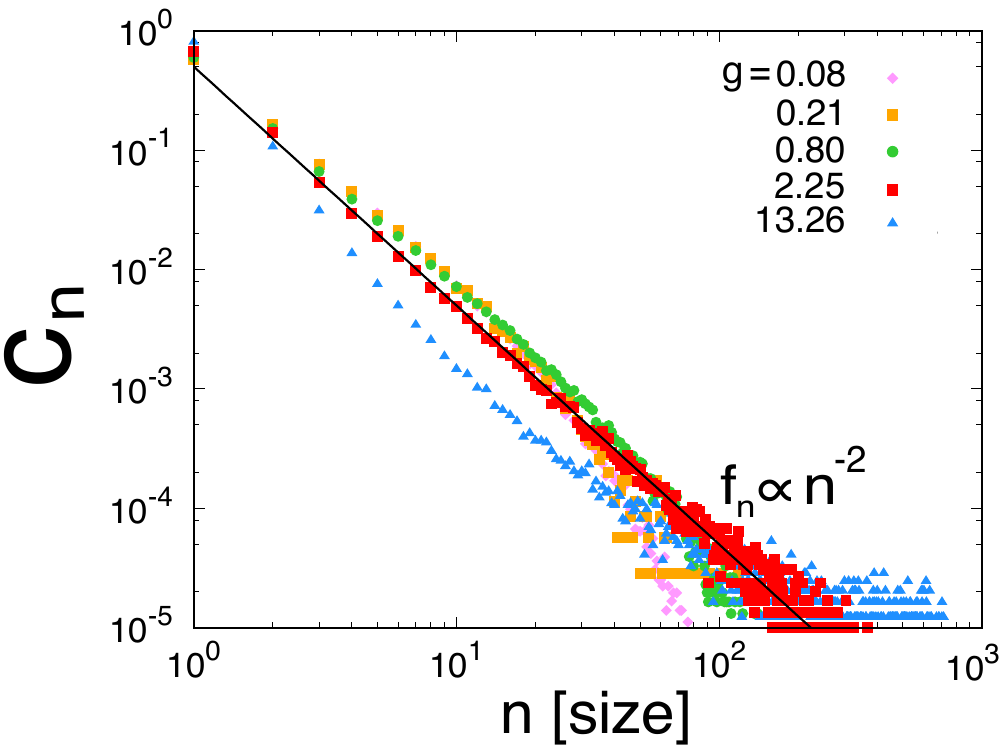} 
\caption{Cluster-size distribution $c_n$ for $\text{Pe}=333.3$  and several $\text{g}$. As $\text{g}$ is increased the distribution changes from an exponential  decay to an algebraic  $1/n^2$ decay for $\text{g}\approx2.25$. At higher  $\text{g}>\text{g}^*\approx2.25$, exponential with a big bump indicating a big cluster. Just in the transition we can appreciate how the  $1/n^2$ decay  fits the data points.}\label{fig:CDF}
\end{figure}

The system of polar ABPs exhibits a transition from a fluid made of a distribution of finite-size clusters to a condensed state wherein the largest cluster occupies a finite fraction of the system.  This large structure is not static but it is constantly changing its shape while particles are exchanged between the dense and the dilute region. In order to study this heterogenous structure we analyze the cluster-size distribution  $c_n$, defined as the distribution function of clusters sorted by sizes $n$. Clusters are defined from a simple overlap criterion: particles with a separation smaller than the range of alignment $R_{\theta}$ are said to share a "bond". A cluster is then the set of all particles that are mutually bonded, and its size is the total number of particles that belong to it.  
In practice, we identify such structural transition through the analysis of $c_n$. 
Below some threshold $\text{g}^*$, the distribution of cluster sizes decays exponentially and broadens as the alignment strength increases, until it eventually becomes  scale-free at  $\text{g}=\text{g}^*$ (see Fig. \ref{fig:CDF}).   Above this value, a macroscopic phase cluster emerges which in the distribution translates into a peaked contribution at large sizes. Thus,  the `microscopic' finite-size clusters regime (mC) is characterized by an exponential distribution of sizes.  The scale-free character $c_n\sim n^{-\alpha}$ of the cluster-size distribution in the ordered phase of the Vicsek model was already noticed in \cite{Huepe2004, Chate2008}, where a power $\alpha\lesssim 2$ was observed in the homogeneous polarized state, while $\alpha\gtrsim2$ closer to the onset of flocking where dense traveling bands are in coexistence with an incoherent background. A similar behavior was also reported for non-aligning self-propelled hard disks in \cite{Levis2014},  with a universal power  $\alpha\approx 1.7$ associated with a percolation transition; and for self-propelled hard rods in \cite{Peruani2006}, also displaying a structural phase transition indicating the emergence of a macroscopic structure. In our model, we find an exponent $\alpha\approx 2$ which weakly depends on the distance to the transition point. 

\begin{figure*}
  \includegraphics[scale=0.6]{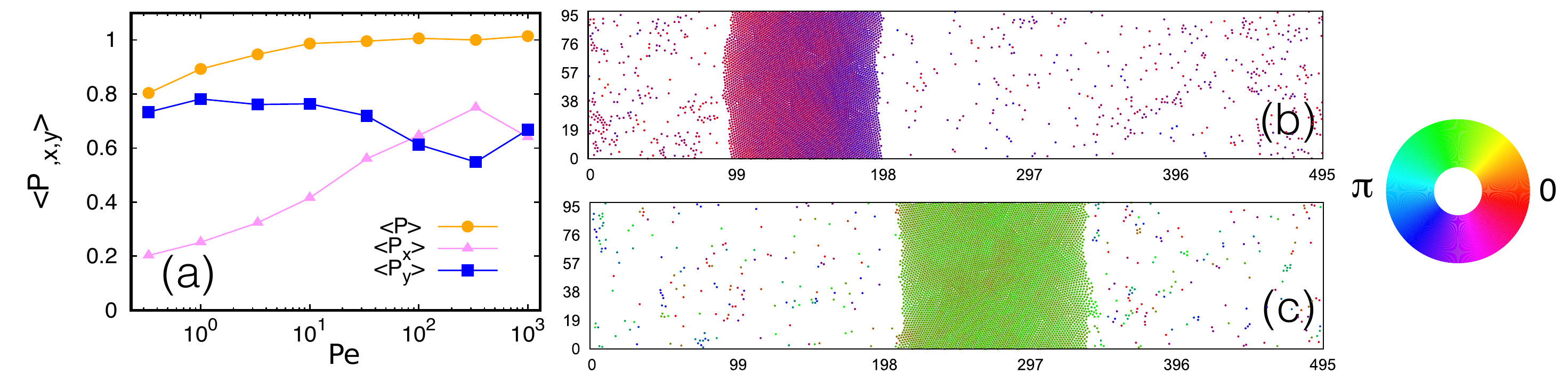} 
  \caption{ (a) The three synchronization order parameter as function of $\text{Pe}$ and constant $\text{g}=10$ (vertical line in the phase diagram of Fig.\ref{fig:PhD}). This configurations have been carried out for a larger system size $N_{x,y}=6N$ and $L_x=5L_y$.
Representative snapshots of a rectangular box ($L_x=498$ , $L_y=98$) with $N=6000$. Corresponding points in the phase diagram; (b) $\text{Pe} = 100$ and $\text{g}=10$, a band structure is observed, and (c) $\text{Pe} =3.33$ and $\text{g}=10$ for which a lane structure is observed. }\label{fig:Pxy}
\end{figure*}

We turn now into the description of the dense structures found in the MC regime. 
Phenomenologically, the recurrence of finding large elongated polarized structures is higher for larger $\text{g}$. 
Strikingly though, at a given value of  $\text{g}$, band and lane structures can be found, and their relative statistical weight is controlled by persistence, as we show below. 
In order to give support to this claim, we run simulations on a rectangular surface( $L_x\times L_y$ with $L_x=5L_y$) with periodic boundary conditions. The choice of this geometry favors the formation of a macroscopic stripe  along the short-axis $\mathbf{e}_y$, highlighting its band- or lane-like nature (see snapshots Fig. \ref{fig:Pxy} and the movies in \cite{SM}).  

In order to identify systematically the band- or lane-like nature of the structure, we define, $P_x$ and $P_y$ as
\begin{equation}
P_x = \left\langle \left |  \frac{1}{N} \sum_{i=1}^N\cos(\theta_i) \right | \right\rangle \, ; \, \, \,  \,  \,  \,  P_y = \left\langle  \left |  \frac{1}{N} \sum_{i=1}^N\sin(\theta_i) \right | \right\rangle\, ,
\end{equation}
where $\left\langle.\right\rangle$ denotes an average over several independent configurations. These observables quantify the net polarization along the longitudinal and transverse direction in the slab geometry. 
We show in Fig. \ref{fig:Pxy}   the behavior of $P$, $P_{x}$ and $P_{y}$ as a function of the persistence of the system  at fixed $\text{g}=10$. 
For  $\text{Pe}\lesssim100$, the longitudinal order parameter is smaller than the transverse one, $P_x <P_y$, confirming that lanes prevail in the small $\text{Pe}$ regime (however, even in this regime, band and lane structures coexist, see \cite{SM}). In contrast, for higher $\text{Pe}$ we find a change in tendency from lanes to bands since $P_y\approx P_x$, showing that the system's structure is characterized by a coexistence of bands and lanes in this regime. 

Why  lanes appear more frequently than bands at small persistence? 
When velocity alignment is much faster than rotational diffusion ($\text{g}\gg \text{Pe}$), particles aggregation is controlled by alignment, following the mechanism depicted in Fig. \ref{aline} (left). 
In this high coupling regime,  rotational diffusion is largely suppressed ($\tau\gg\tau_K$).
Once a particle aligns with the cluster its orientation gets kinetically frozen, due to the separation of time scales between $\tau$ and $\tau_K$. 
In this regime, a coherently moving cluster grows by the aggregation of particles along their direction of motion, favoring a lane-like structure  (see Fig. \ref{fig:laneband} and Movie 1 in the Supplementary Information \cite{SM}). 


As $\text{Pe}$ increases, fluctuations are more severe and the orientation of the particles cannot be considered as frozen anymore since  
the persistence time $\tau$ can be of the order of magnitude of $\tau_K$. Therefore, a reorientation event might make a particle leave the surface of the cluster if its diffusion time is short enough compared to $\tau_K$. In the absence of interactions, an Active Brownian Particle has a diffusivity $D\propto \text{Pe}^2$; so the higher $\text{Pe}$, the shorter is $\tau$ and the easiest is for a particle to leave a lane.
This can be heuristically described in terms of an effective temperature, as put forward in the context of co-fluent tissue models in \cite{Giavazzi2017}. Following \cite{Giavazzi2017}, if one considers the macroscopic coherent structures as a solid - where particles inside are caged by their neighbors - then, one can argue that fluctuations are well captured by an effective temperature $T^K_{\text{eff}}\propto \text{Pe}^2/\text{g}$. 
This explains why lanes have a tendency to become less stable as $\text{Pe}$ increases. 
Still, this does not explain why the typical band structures of the Vicsek model take over as $\text{Pe}$ increases. 

At high $\text{Pe}$, fluctuations in the orientations are large, so the orientational degrees of freedom of the particles in the macro-cluster cannot be considered as frozen anymore. In this regime,  we observe bands and lanes structures. As illustrated  in Fig.  \ref{fig:laneband}, it is harder for a particle in the surface of a band to leave the cluster than for a particle in the surface of a lane. 
If a particle in the surface of a lane picks a random orientation pointing outwards, and the persistence of its motion is high enough, it will interact with less neighbors as its persistent motion proceeds and eventually leave the cluster. The way particles are structured in a band makes it more robust against escape processes as the one just described. 
As illustrated  in Fig.  \ref{fig:laneband}, a random re-orientation of a particle in the surface of a band cannot lead to a reduction of the number of neighbors. However, because of the persistence of the particles motion and the small inter-particle spacing, random fluctuations of the orientation induce further hard-core collisions (see overlapping disks in Fig. \ref{fig:laneband}). This enhancement of collisions in the bulk and forefront of a traveling (dense) band introduces extra fluctuations which might destabilize it, resulting in the coexistence (or bistability) of lanes and bands observed in the high $\text{Pe}$ regime (see Fig.  \ref{fig:Pxy} and Movie 2 and 3 in \cite{SM}). A random fluctuation of the orientation of a particle may leave an empty space that, in turn, may help another particle to escape. This is a simple illustration of the intricate coupling between density and polarization fluctuations at short length scales due to the presence of volume exclusion. 
The competition between the stability of the band from the point of view of the alignment interactions, and the enhancement of collisions due to Active Brownian motion -  absent in the lane structure - is introduced by excluded volume effects and, as such, it is absent in traditional Vicsek-like models of point-like agents. 



\begin{figure}
  \includegraphics[scale=0.25]{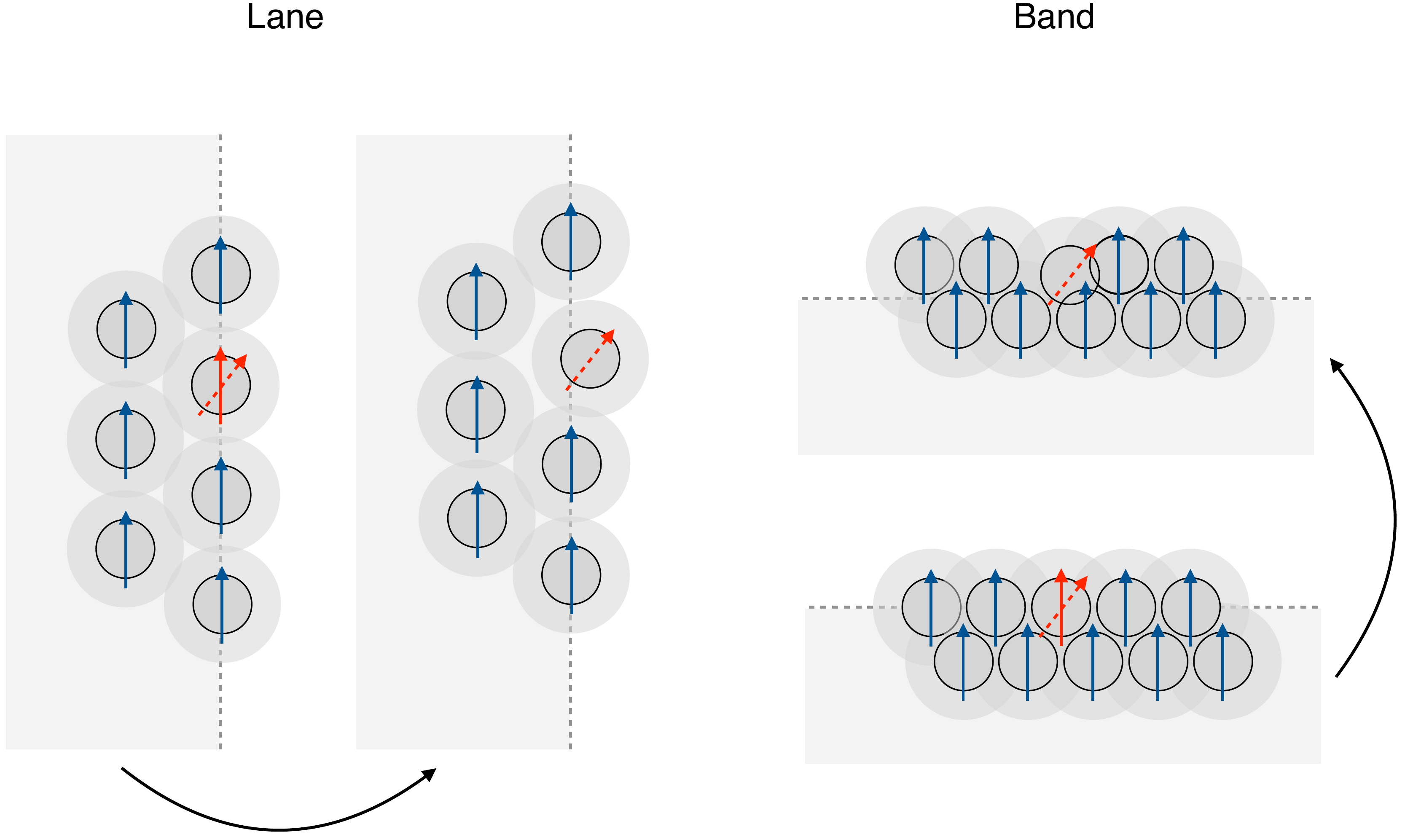}
\caption{Illustration of the mechanism behind the stability of lanes and bands. In the low persistence regime the alignment time scale is much smaller that the reorientation time, thus freezing the structure of the lane. At higher persistences, fluctuations in the orientation, represented by the discontinuous red arrow, are stronger, rendering the lane structure unstable and favoring the formation of a band. As shown in the right figure, a band structure is more robust against fluctuations in the orientation than a lane, the relative distance between particles shrinks as persistence increases and fluctuations in the orientation might foster inter-particle collisions. }\label{fig:laneband}
\end{figure}

\begin{figure}
  \includegraphics[scale=0.85]{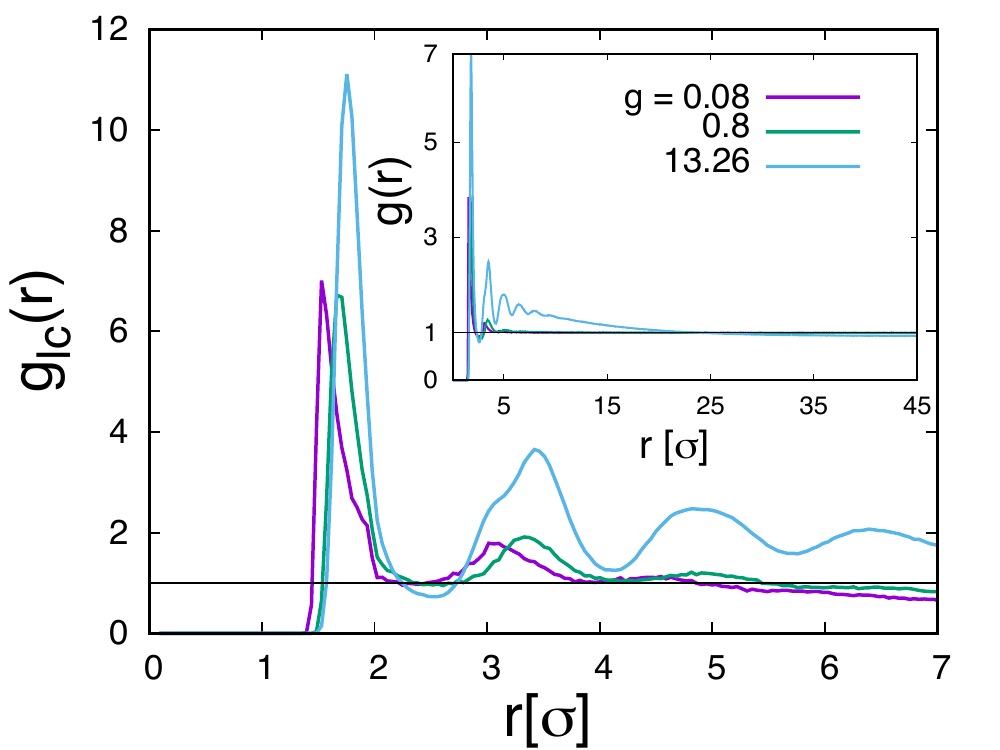}
  \includegraphics[scale=0.85]{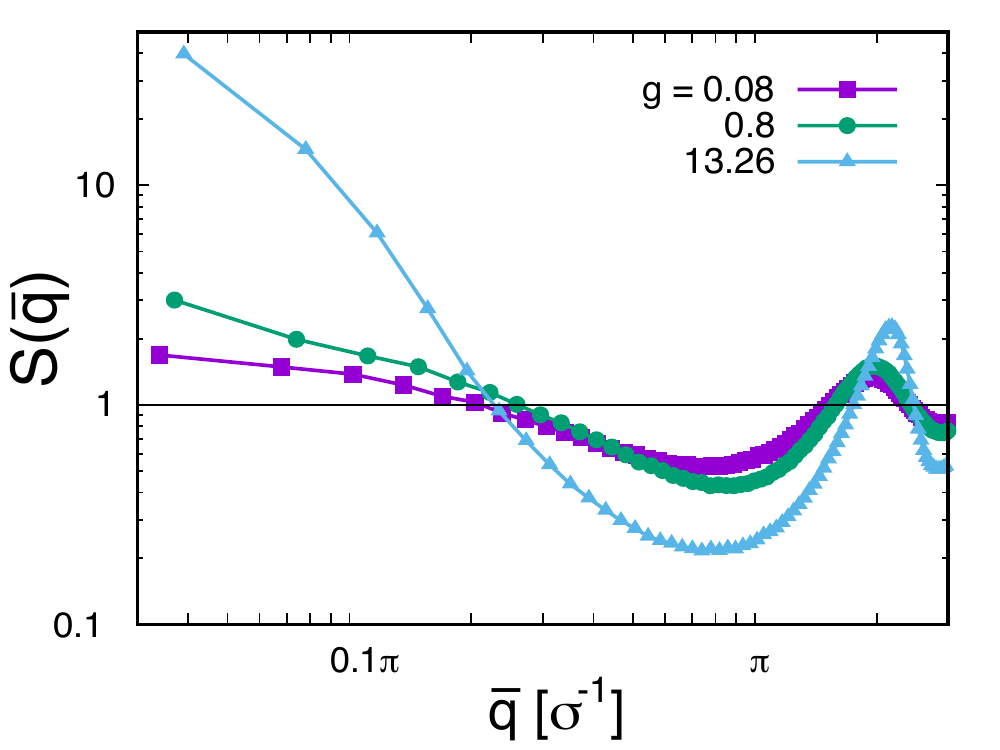}
\caption{The radial distribution function of the largest cluster $g_{lc}(r)$ for $\text{Pe}=333.3$ and several $\text{g}$.   Inset: The averaged radial distribution functions $g(r)$ in the interval $r \in [0,L/2]$ for $\text{Pe}=333.3$  and several $\text{g}$ show the decay of the tails approaching to $g(r)=1$ at large distances. 
The structure factor in log-log scale for $\text{Pe}=333.3$ and different values of $\text{g}$. The functions have been normalized through the equilibrium distance $r_0$ obtained from the first peak of $g(r)$. }\label{fig:gr}
\end{figure}

We characterize the main structural properties of the system by analyzing the behavior of the radial distribution function, 
\begin{equation}
g({r})= \frac{1}{N} \left \langle \sum_{j\neq i}\sum_i \delta (r-|\mathbf{r}_i-\mathbf{r}_j|)\right \rangle
\end{equation} 
 and the structure factor,
 \begin{equation}
S(q)= \frac{1}{N}\left \langle  \left | \sum_{i=1}^N e^{i\mathbf{q}\cdot\mathbf{r}_i}  \right | ^2 \right \rangle
\end{equation}
These quantities are shown in Fig. \ref{fig:gr}. The radial distribution function together with the contribution from the particles that belong to the largest cluster, $g_{lc}(r)$, as function of several $\text{g}$ and fixed $\text{Pe}$.
When increasing $\text{g}$ at fixed $\text{Pe}$, the height of the peaks of the radial distribution function increases, reflecting an increasing  degree of order of the structure. 
It is also noteworthy that increasing $\text{g}$ at fixed Pe also produces a shift in the peaks to larger distances. This shift provides a clear evidence that highly coupled particles keep larger relative distances since their orientation gets effectively frozen as soon as they are in the macroscopic cluster, within the interaction range $R_{\theta}$. As shown in Fig. \ref{aline} while two interacting particles are approaching each other they experience a rapid orientation along the direction of its neighbors in a time scale $t \propto \tau_K$. Once their orientation is locked with the one of their neighbors, they will move parallel and the possibility of getting closer is drastically reduced.

Analogously, $S(q)$ shows a maximum at $q \simeq 2\pi/r_0$, where the typical distance between two neighboring disks  is $r_0 \simeq \sigma$. In Fig. \ref{fig:gr}, we have represented $S(q)$ as a function of the scaled quantity $\bar{q}=q \cdot r_0$.
Now, the peak in $2\pi$ reflects the influence of the potential repulsion, which controls the short-ranged structure of a simple liquid. While increasing $\text{g}$, $S(q)$ strongly increases at low-$q$ and  at $\bar{q}=2\pi$. This emerging density fluctuations at low-$q$ directly reveal the presence of the clusters observed in real space, which produce an inhomogeneous structure over a much larger length scale than simple fluids. Nonetheless, the increase  at $2\pi$ points out that clusters are denser structures than the corresponding ones for a passive system.

\section{Discussion}\label{sec:conclusion}

We have analyzed the impact of excluded volume interactions in a model undergoing a flocking transition. We have shown that the competition between self-propulsion, alignment and excluded volume gives rise to richer non-equilibrium structures than the Vicsek model and the ABP model, both recovered as special limits of our model. We have established a phase diagram summarizing the impact of self-propulsion and velocity alignment in the selection of different structures:  isotropic and polarized micro-clusters, and macroscopic structures which are generically made of a statistical mixture of lanes and bands. In particular, we shed light into the crucial role played by excluded volume effects in the selection of the macroscopic traveling structure, and show how the persistence of the particles motion controls its lane- or band-like character.  Similar patterns have been observed in previous models which include density-dependent motility \cite{Farrell2012} and experiments on actin filaments \cite{Schaller2010, Schaller2011polar, Kohler2011}, showing that crowding has a generic impact on  structure formation in active matter. The competition between short-range repulsion and velocity alignment yields a complex clustering mechanism which is controlled by the coupling between  density and polarization fluctuations at short length scales.   Understanding the stability of bands and lanes at different persistences from a theoretical coarse-grained hydrodynamic  description remains an open challenge calling for further developments.  

\section*{Acknowledgments}
A.M.-G. acknowledges received funding from the European Union's Horizon 2020 research and innovation programme under grant agreement No 674979-NANOTRANS.
D.L. acknowledges received funding from the European Union's Horizon 2020 research and innovation programme under the Marie Sklodowska-Curie (IF) grant agreement No 657517. 
A.D.-G. acknowledges financial support from MINECO under projects No. FIS2012-38266-C02-02 and No. FIS2015-71582-C2-2-P (MINECO/FEDER); and Generalitat de Catalunya under the grant No. 2014SGR608. 
I.P. acknowledges MINECO and DURSI for financial
support under projects FIS2015-67837- P and 2014SGR-922,
respectively. 

\bibliographystyle{apsrev}
\bibliography{AITOR}

\end{document}